\documentclass[prD,superscriptaddress,twocolumn,showpacs,nofootinbib,amsmath,amssymb]{revtex4-1}
\usepackage{graphicx,bm,adjustbox,mathrsfs}
\usepackage{color,ulem}

\begin{document}
\pagenumbering{arabic}

% Authors, title and abstract
\title{Constraints on MeV dark matter using neutrino detectors and their
implication for the 21-cm results}

\author{Niki Klop}
\email{l.b.klop@uva.nl}
\affiliation{GRAPPA Institute, University of Amsterdam, 1098 XH
Amsterdam, The Netherlands}
\author{Shin'ichiro Ando}
%\email{s.ando@uva.nl}
\affiliation{GRAPPA Institute, University of Amsterdam, 1098 XH
Amsterdam, The Netherlands}
\affiliation{Kavli Institute for the Physics and Mathematics of the
Universe (Kavli IPMU, WPI), Todai Institutes for Advanced Study,
University of Tokyo, Kashiwa, Chiba 277-8583, Japan}
\begin{abstract}
The recent results of the EDGES collaboration indicate that during the
 era of reionization, the primordial gas was much colder than
 expected. The cooling of the gas could be explained by interactions
 between dark matter (DM) and particles in the primordial
 gas. Constraints from cosmology and particle experiments indicate that
 this DM should be light ($\sim$10--80 MeV), carry a small charge
 ($\epsilon\sim 10^{-6}$--$10^{-4}$), and only make up a small
 fraction of the total amount of DM. Several constraints on the DM
 parameter space have already been made. We explore the yet
 unconstrained region in the case that the milli-charged DM makes up for
 $\sim$2\% of the total dark matter, through the scenario in which this
 DM annihilates only into mu and tau neutrinos. We set upper limits on
 the annihilation cross section using the Super-Kamiokande data, and
 predict the limits that could be obtained through Hyper-Kamiokande,
 JUNO and DUNE. We find that data from Super-Kamiokande is not yet able to constrain this model, but future experiments might be.
We furthermore explore DM annihilation into solely neutrinos in general, giving an update of the current limits, and
 predict the limits that could be placed with future experiments.
\end{abstract}
\date{\today}
\maketitle

% Chapters
\section{Introduction}
\label{sec:intro}

Early stars are expected to have imprinted their
evidence in the cosmic microwave background (CMB). 
Their ultraviolet light hit the primordial hydrogen gas, resulting in
emission at the 21 cm line. 
As a response to this, the absorption from CMB photons by the primordial
gas caused a spectral signal in the CMB that we should be able to
observe today.
The recent 21-cm results of the EDGES
collaboration~\cite{Bowman:2018yin} show an absorption profile that is
consistent with the expected effect induced by the early stars, although
showing an amplitude twice as large as predicted. 
This result implies that the temperature of the primordial gas was much
lower than expected, or that the temperature of the background radiation
was higher than expected. 

A possible explanation consistent with the observed results is the
cooling of the gas due to interactions with dark matter
(DM)~\cite{Bowman:2018yin,Barkana:2018lgd,Fialkov:2018xre,Tashiro:2014tsa,Munoz:2015bca}, which is causing
a lot of excitement in the field. The possibility of such a DM interaction is studied in Refs.~\cite{Munoz:2018pzp,
Berlin:2018sjs,Kovetz:2018zan}, in which multiple constraints are put on the nature of
the responsible DM. 
Using data from a variety of experiments, it is found that most of the
parameter space that is consistent with the 21-cm observations is ruled
out~\cite{Berlin:2018sjs}. 
The DM responsible for the cooling could only make up for a small
fraction of the total DM, $\sim$0.3--2\%, and their mass lies in the
range of $\sim$10--80~MeV. 
Furthermore, the DM should carry a small electric charge in the order of
$\epsilon\sim 10^{-6}$--$10^{-4}$.

However, assuming that DM interacts with baryons mediated by only
photons produces too much DM through thermal freeze-out mechanism. 
In order to circumvent this issue, DM must have at least one more
interaction channel with the standard model particles.
The simplest possilibity that has not been ruled out yet is that the DM
interacts with lepton number $L_{\mu} - L_{\tau}$ via either a scalar or
vector mediator~\cite{Berlin:2018sjs}.
This model is hard to constrain with lab experiments because DM does not
interact with electrons, and especially for DM lighter than muons (as
it is of main interest here), DM can annihilate only into muon and tau
neutrinos.

We investigate this scenario, exploring the yet unconstrained parameter
space in the energy range of $\sim$10--100~MeV. 
Through flavor mixing, $\nu_e$ and $\bar\nu_e$ have been generated when the
neutrinos reach the Earth, which makes it possible for detectors such as
Super-Kamiokande (SK)~\cite{Bays:2011si} to detect them through charged-current
interations.
These neutrinos will show a very specific spectral feature; for example,
in the simplest model investigated in Ref.~\cite{Berlin:2018sjs}, DM
annihilation will produce a neutrino line at its mass ($\chi\chi \to
\nu\bar\nu$).
The energy range of $\sim$10--100~MeV, where there are solar, reactor,
and atmospheric neutrino backgrounds as well as cosmic ray muons, has
been studied well especially for detecting the diffuse supernova
neutrino background~\cite{Ando:2004hc, Beacom:2010kk,
Lunardini:2010ab}.

We obtain upper limits on the annihilation cross section of this DM in
the case that it makes up $\sim$2\% of the total DM, the model that
could explain the EDGES result, using the several years of SK data. 
We also predict the upper limits that could be obtained by future
experiments, Hyper-Kamiokande (HK)~\cite{Abe:2018uyc}, Deep Underground Neutrino Experiment (DUNE)~\cite{Strait:2015aku} and Jiangmen Underground Neutrino Observatory (JUNO)~\cite{Antonelli:2017uhq}. 
A DM model like this, in which DM only annihilates into neutrinos, but
making up for the entire amount of DM, has been studied
before~\cite{PalomaresRuiz:2007eu,Campo:2018dfh}, obtaining upper limits on the cross
section using the SK data. 
We also obtain updated upper limits for this scenario. 

The paper is organised as follows.
In Sec.~\ref{sec:theory}, we discuss milli-charged DM and the DM scenario
we investigate.
In Sec.~\ref{sec:DMFlux}, we determine the neutrino flux coming from the
annihilation of this milli-charged DM, while in Sec.~\ref{sec:DMFlux}, we
explain the analysis we perform. 
In Secs.~\ref{sec:results} and \ref{sec:conclusion}, we discuss our
results and conclusions respectively.

\section{theory}\label{sec:theory}

\subsection{Milli-charged DM}

There are some requirements for the DM properties that need to hold in
order to be responsible for the extensive cooling of the primordial
hydrogen gas~\cite{Berlin:2018sjs}: Due to the equipartition theorem,
the DM particles should be relatively light. 
Furthermore, models in which the cross section for dark matter scatterings with gas
is independent of the velocity can already be ruled out by constraints
from observations of the CMB. 

To fulfil these requirements, the mediator of the dark matter-baryon
interactions should be lighter than the temperature of the gas at $z\sim
17$. 
New light mediators in the mass range required to explain the EDGES
result are ruled out~\cite{Adelberger:2006dh,Kapner:2006si}, and
their contribution to the radiation part of the energy density would
exceed the current constraints \cite{Ade:2015xua}. 
When the DM carries a small electric charge, it could couple to the
photon. 
Reference~\cite{Munoz:2018pzp} finds that, in order to cool the gas
sufficiently, the following condition for the electric charge should be
fulfilled:
\begin{equation}
\epsilon \approx 1.7 \times 10^{-4} \left( \frac{m_{\chi}}{300 \, \mathrm{ MeV}}\right) \left(\frac{10^{-2}}{f_{\chi}}\right)^{3/4},
\end{equation}
where $\epsilon \equiv e_{\chi}/e$ is the electric charge of the
milli-charged DM particle, $m_{\chi}$ is its mass and $f_{\chi}$
is its mass fraction of the total DM. 
The existence of milli-charged DM is already constrained by multiple
experiments and astrophysical data, leaving only a small open window in
its possible parameter space, with a mass $m_{\chi}$ of
$\sim$10--80~MeV, and a total DM fraction of $f_{\chi} \sim
0.003$--0.02~\cite{Berlin:2018sjs}.

However, the annihilation of milli-charged DM particles through the
exchange of a photon is not sufficient to yield the desired energy
density for the particle, $f_{\mathrm{DM}}\Omega_{\mathrm{CDM}}$. 
Therefore, some additional annihilation needs to take place through a
new mediator.
Here we specifically study the case of vector mediator $V$, while the
results for the scalar mediator are essentially the same.
Refeference~\cite{Berlin:2018sjs} finds that annihilation through the
new vector mediator $V$ into standard model fermions is excluded if it
couples to all flavors. 
This leads to the consideration of annihilation into mainly neutrinos. 
When the new vector $V$ is related to the gauge group
$U(1)_{L_{\mu}-L_{\tau}}$, only coupling to muons, taus, muon-neutrinos
and tau-neutrinos is possible. 
Since $V$ does not couple to electrons, there are not yet many
constraints from experiments. 
The annihilation cross section to any neutrino flavor for such a model
is given by
\begin{equation}
\langle \sigma v \rangle =
 \frac{g^2_{\nu}g^2_{\chi}m^2_{\chi}\kappa}{2\pi(4m^2_{\chi}-m^2_V)^2},
\end{equation}
where $g_{\nu}$ and $g_{\chi}$ are the gauge
coupling constants of the neutrino and DM particle, respectively, and
$\kappa = 1$ $(v^2/6)$ for fermion (scalar) DM.
We put constraints on this model by evaluating the SK data, and make
predictions for some future experiments.

Beside this, we also consider DM annihilation into neutrinos in a
broader sense. 
If dark matter only annihilates into neutrinos, this would be harder to
detect than the cases where gamma rays are produced. 
The limits on the annihilation cross section in this case will therefore
be the most conservative ones, and therefore interesting to
investigate. 
We update the limits obtained by Ref.~\cite{PalomaresRuiz:2007eu},
calculating the limits both in the case that DM annihilates to all three
neutrino flavors, as in the case that it only annihilates into muon and
tau neutrinos as discussed above.

\section{Neutrino flux from dark matter annihilation}\label{sec:DMFlux}

The final flavor ratio on Earth for pure $\nu_{\mu}$ and $\nu_{\tau}$
channels is $1:2:2$. 
When Galactic DM annihilates into $\nu_\mu$ and $\nu_\tau$, the expected
monochromatic flux of electron (anti-)neutrinos at Earth will therefore
be given by
\begin{equation}\label{eq:flux}
\frac{\mathrm{d}\phi}{\mathrm{d}E_{\nu}} = \frac{\langle \sigma v\rangle}{2}\mathcal{J}_{avg}\frac{R_{sc}\rho_0^2f_\chi^2}{m_{\chi}^2}\frac{1}{5}\delta(E_\nu - m_\chi),
\end{equation}
in the case of Majorana DM, where $\sigma$ is the annihilation cross
section, $m_\chi$ is the mass of the DM particle, $E_\nu$ is the
neutrino energy, $\mathcal{J}_{avg}$ is the angular-averaged
``$J$-factor'' of the Milky Way, for which we use the canonical value
$\mathcal{J}_{avg}=5$~\cite{Yuksel:2007ac}, $R_{sc}=8.5$~kpc is the scale radius of the Milky Way, and
$\rho_0 = 0.3$~GeV~cm$^{-3}$ is the DM density at the scale radius.
To retrieve the electron (anti-)neutrino flux for Dirac DM,
Eq.~(\ref{eq:flux}) has to be divided by 2.

In the thermal freeze-out scenario, the annihilation cross section at
freeze-out required to leave the correct relic abundance of MeV DM is given
by
\begin{equation}\label{eq:sigmav}
 \langle\sigma v\rangle = \frac{5\times
  10^{-27}~\mathrm{cm^3~s^{-1}}}{\Omega_{\chi}h^2},
\end{equation}
for Majorana fermion DM, and is twice as large for Dirac fermion
DM for masses below GeV~\cite{Steigman:2012nb}.
Since $\Omega_{\chi}h^2 \approx 0.1 f_\chi$, the targeted annihilation cross
section is $\langle \sigma v \rangle \approx 2.5 \times
10^{-24} (f_\chi/0.02)^{-1}~\mathrm{cm^3~s^{-1}}$ and $5 \times
10^{-24} (f_\chi/0.02)^{-1}~\mathrm{cm^3~s^{-1}}$ for Majorana and Dirac
DM respectively.

\begin{figure}
\includegraphics[width=8.5cm]{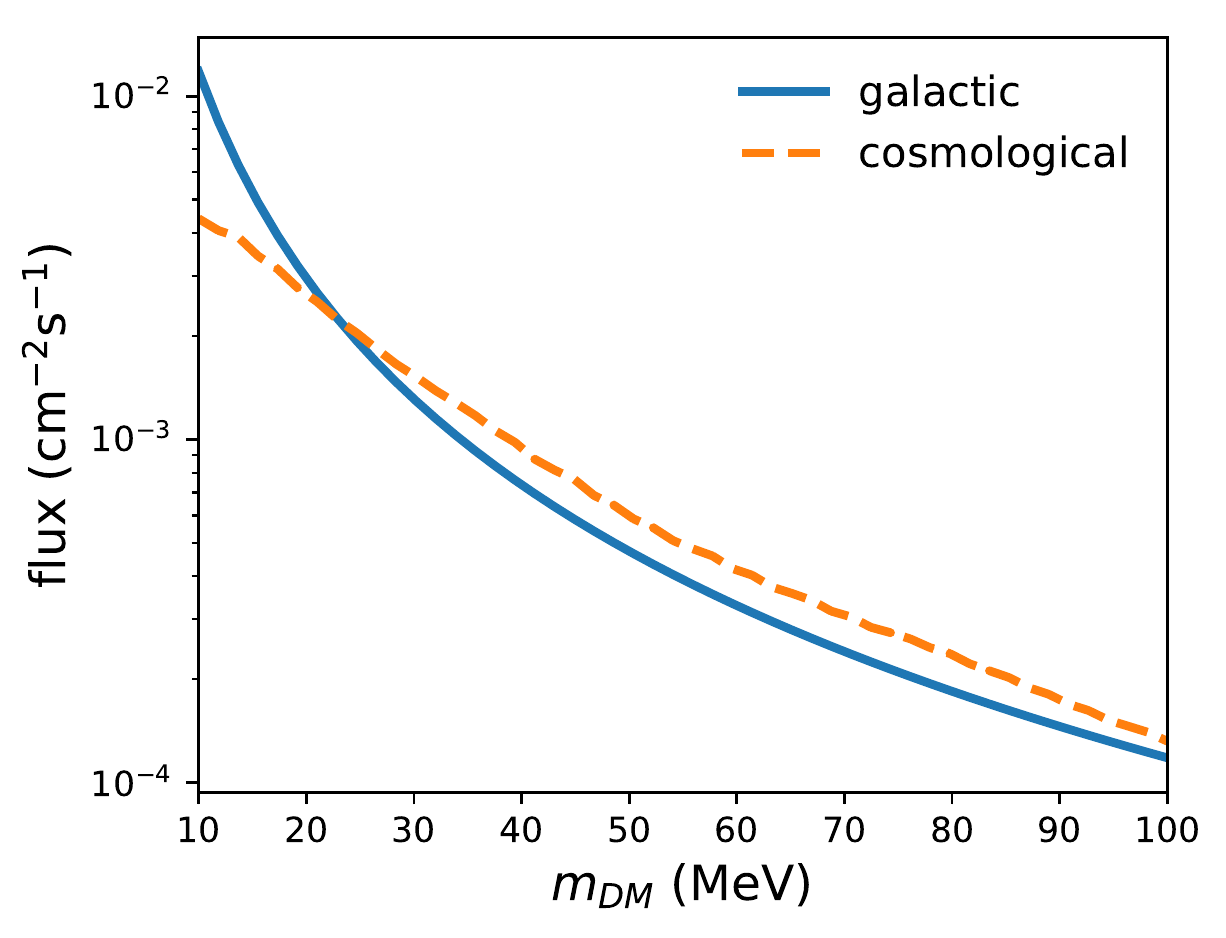} 
\caption{\small The total integrated electron neutrino flux at Earth of both the cosmological and the Galactic contribution as a function of the DM mass, in the case of $2\%$ Dirac DM annihilating into only muon and tau neutrinos.}\label{fig:flux}
\end{figure}

Besides the Galactic neutrino flux, we also take into account the
contribution to the flux coming from DM annihilations outside our
galaxy.
We adopt the calculation of Ref.~\cite{Ando:2013ff} with the most recent 
model of substructure boost~\cite{Hiroshima:2018kfv}.
This cosmological neutrino flux is of the same order of magnitude as the
Galactic contribution, but non-monochromatic due to its redshift. 
In Fig.~\ref{fig:flux}, the total integrated flux of both the
cosmological and the Galactic contribution are shown as a function of
the DM mass.

\section{The analysis}\label{sec:analysis}

We set upper limits on the DM annihilation cross section using the
latest SK data~\cite{Bays:2011si}, and predict the upper limits that
could be obtained by the future experiments,
Hyper-Kamiokande~\cite{Abe:2018uyc}, DUNE~\cite{Strait:2015aku} and JUNO~\cite{Antonelli:2017uhq}.
We use the SK data from the first three data periods~\cite{Bays:2011si},
which contains 2853 days of data taking in total, in the energy range of
16--88~MeV, considering 18 bins with a width of 4~MeV.
The expected number of events at the detector coming from DM
annihilation is calculated through
\begin{equation}
N_{\mathrm{events}} = \sigma_{\mathrm{det}}\phi N_{\mathrm{target}}\epsilon_{\mathrm{det}}t,
\end{equation}
where $\sigma_{\mathrm{det}}$ is the detection cross section, $\phi$ is
the neutrino flux, $N_{\mathrm{target}}$ is the number of target
particles in the detector, $\epsilon_{\mathrm{det}}$ is the efficiency
of the detector which we get from~\cite{Bays:2011si}, and $t$ is the
exposure time. 
SK is a 22.5 kton water Cherenkov detector~\cite{Bays:2011si},
detecting neutrinos through the measurement of Cherenkov radiation from
relativistic electrons and positrons.
The relevant detection channels in our energy range are inverse beta
decay ($\bar{\nu}_{e}+p \rightarrow e^+ +n$), and the absorption of
$\nu_e$ and $\bar{\nu}_e$ by Oxygen in charged current interactions
($\overset{(-)}{\nu_e} + ^{16}\mathrm{O} \rightarrow e^{\pm} +
\mathrm{X})$. 
The energies of electrons and positrons produced by these interactions
are $E_e = E_\nu - 1.3$~MeV ($\bar\nu_e p$), $E_\nu - 15.4$~MeV ($\nu_e
O$), and $E_\nu - 11.4$~MeV ($\bar\nu_e O$).
The cross sections for these detection channels are taken from
Refs.~\cite{Strumia:2003zx, Skadhauge:2006su}. 
To correct for the energy resolution of the experiment, we smear the
expected electron (positron) spectrum with a Gaussian function, using an
energy resolution of width
\begin{equation}\label{eq:resSK}
\sigma =0.40~\mathrm{MeV} \sqrt{E/\mathrm{MeV}}+0.03E,
\end{equation}
that we take from Ref.~\cite{PalomaresRuiz:2007eu}. 
We perform a $\chi^2$ analysis of the expected number of events compared
to the data, and calculate the upper limit at the 90\% confidence level. 
We consider four different backgrounds coming from atmospheric
neutrinos, that we also take from Ref.~\cite{Bays:2011si}. 
This background data is taken from the first running phase of SK, SK-I. 
We rescale it to the entire exposure time that we consider of 2853
days.

The future experiments that we consider show a lot of improvement in
several ways. 
The invisible muon background originating from $\nu_\mu/\bar{\nu}_\mu$
charged current events, which is the biggest background in SK below
$\sim$40~MeV, might be significantly decreased in measurements of future
water Cherenkov detectors by adding Gadolinium~\cite{Beacom:2003nk}.
In our analysis for Hyper-Kamiokande, we assume a reduction of this
background by 80\%. In the case of JUNO, this background is removed in total through the implementation of an extra system for cosmic muon detection and background reduction \cite{Lu:2017ciz}.
In the case of DUNE, this background is removed (e.g., see
\cite{Cocco:2004ac}).
For our prediction, we consider an exposure time of 3000 days for each
detector.
We use the same background data as in SK, rescaling it to the right
exposure time and the size of the specific experiment. 
In the case of Hyper-Kamiokande, a 520 kton upgrade
of SK ~\cite{Abe:2018uyc} with a fiduciul volume of 374 kton, the same energy resolution is used
[Eq.~(\ref{eq:resSK})]. 

DUNE is a 46.4-kton liquid argon detector~\cite{Kudryavtsev:2016ybl}.
The relevant detection channels are the detection of electron
(anti-)neutrinos through charged current interactions
($\nu_{e}/\bar{\nu}_{e}+^{40}\mathrm{Ar} \rightarrow e^-/e^+ +
\mathrm{A}' + nN$)~\cite{Skadhauge:2006su}, where $nN$ are the emitted
nuclei and $\mathrm{A}'$ is the remaining nucleus. 
JUNO is a 20-kton lab based liquid scintillator~\cite{Antonelli:2017uhq}.
The detection channels relevant for JUNO are again inverse beta decay
($\bar{\nu}_{e}+p \rightarrow e^+ +n$), and the capture of electron
(anti-)neutrinos on $^{12}\mathrm{C}$ in charged current interactions
($\bar{\nu}_e + {^{12}\mathrm{C}} \rightarrow {^{12}\mathrm{B}} +e^+ $/
$\nu_e + {^{12}\mathrm{C}} \rightarrow
{^{12}\mathrm{N}}+e^-$)~\cite{Skadhauge:2006su}.
The energy resolutions for both JUNO and DUNE are significant better
than the water Cherenkov detectors.
For DUNE, the energy resolution is given by~\cite{DeRomeri:2016qwo}
\begin{equation}
\sigma= 0.025~{\rm MeV}\sqrt{E/{\rm MeV}} + 0.060E.
\end{equation}
For JUNO, the energy resolution we use is given by~\cite{An:2015jdp}
\begin{equation}\label{eq:resJUNO}
\sigma =0.03~{\rm MeV} \sqrt{E/{\rm MeV}}.
\end{equation}

To predict the upper limit at the 90$\%$ confidence level, we use the
python tool {\tt swordfish}~\cite{Edwards:2017kqw, Edwards:2017mnf} that
can, among others, predict upper limits based on Poisson statistics,
resulting in approximately the mean value of the results that one would
get performing a Monte Carlo simulation.
Besides this, in the case of Hyper-Kamiokande, we explicitly perform a
Monte Carlo simulation to obtain the full scope of possible values. 
We furthermore consider the reach of a hypothetical experiment with the
size of Hyper-Kamiokande and the specifications of JUNO, which would
result in a very strong experiment.

We obtain the upper limits for several cases. 
Besides the case of a milli-charged Dirac DM particle responsible for
$2\%$ of the total DM, annihilating into only muon and tau neutrinos, we
also consider the situation where the 100\% of DM only annihilates into
neutrinos. 
In this situation we consider two subcases. 
In the first case, the DM has the same properties as in the
2\%-situation. 
In the second case, we consider Majorana DM, annihilating into all three
neutrino flavors. 
Since in this case the expected flavor ratio at Earth is $1:1:1$, the
expected neutrino flux is given by
\begin{equation}\label{eq:flux2}
\frac{\mathrm{d}\phi}{\mathrm{d}E_{\nu}} = \frac{\langle \sigma v\rangle}{2}\mathcal{J}_{avg}\frac{R_{sc}\rho_0^2}{m_{\chi}^2}\frac{1}{3}\delta(E_\nu - m_\chi).
\end{equation}
The latter case is similar to ordinary WIMP DM, except that the
neutrino-only restriction makes it harder to detect. 
Therefore, the upper limits obtained in this situation will be the most
conservative constraints for MeV WIMP DM annihilation.

\section{Results}\label{sec:results}

\begin{figure}
 \begin{center}
  \includegraphics[width=8.5cm]{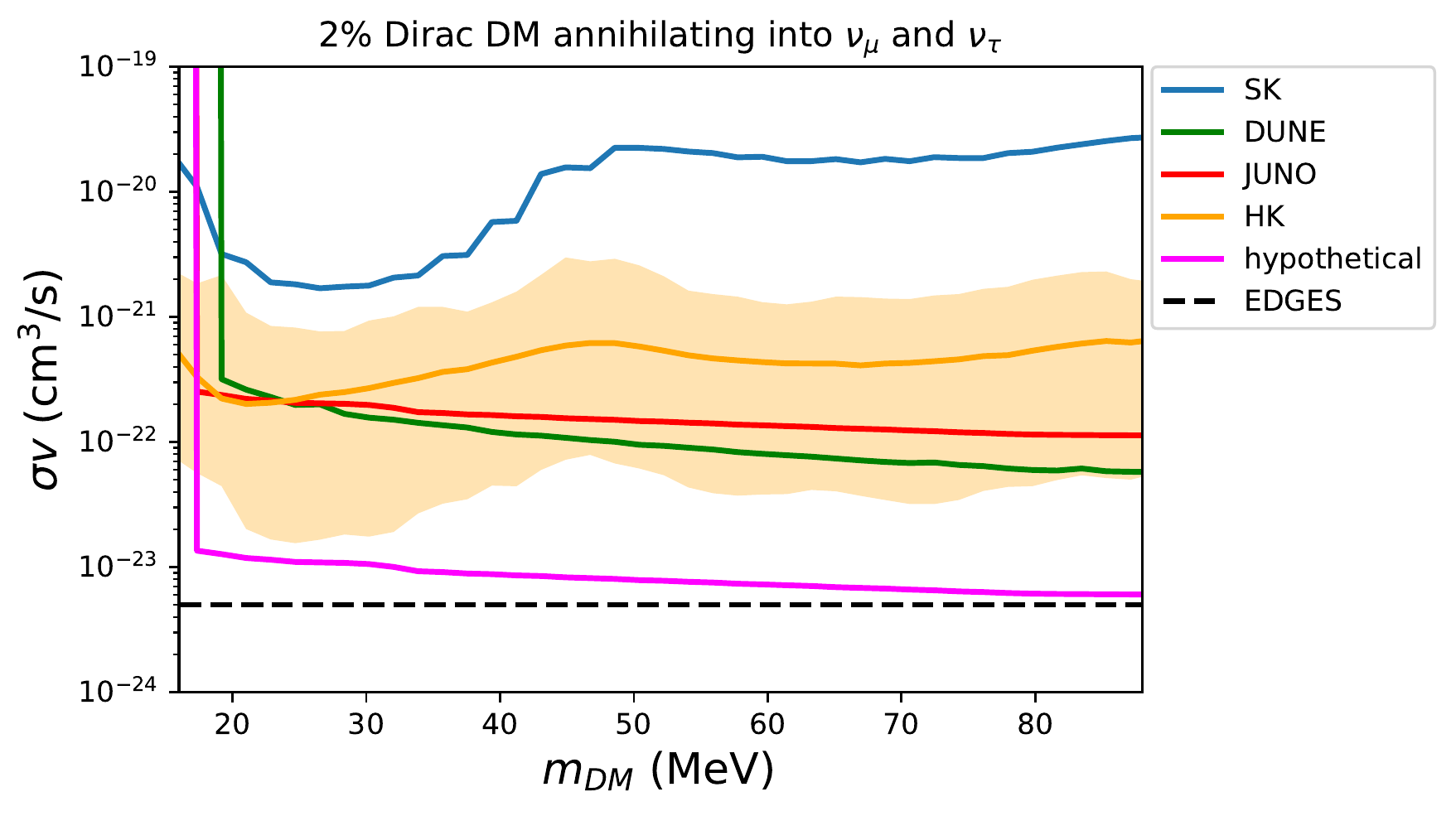} 
  \includegraphics[width=8.5cm]{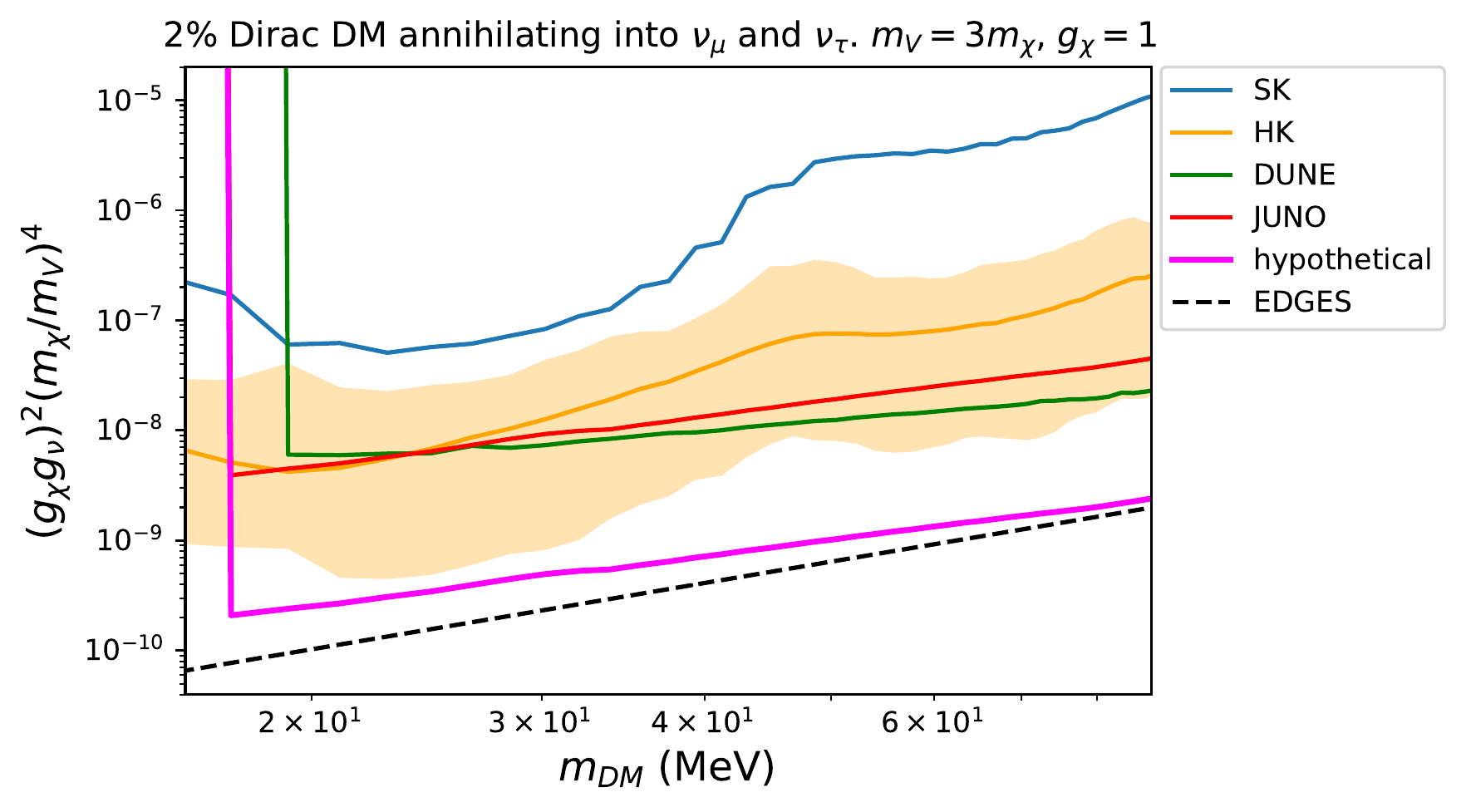}
  \caption{ Upper limits on the annihilation cross section $\langle
  \sigma v\rangle$ of milli-charged Dirac DM into only muon and tau
  neutrinos, making up 2$\%$ of the total DM, as a function of the DM
  mass (upper panel). The lower panel shows the annihilation cross
  section in terms of the coupling constants, the DM mass, and the mass
  of the new mediator $V$. The black dashed line is the cross section
  induced by the EDGES results, in case of Dirac DM. The blue line comes
  from the analysis of 2853 days of SK data. The other lines are
  predictions using Swordfish~\cite{Edwards:2017kqw,
  Edwards:2017mnf}. The orange band shows the region between the minimum
  and maximum upper limit values predicted for the Hyper-Kamiokande
  through a Monte Carlo simulation.}\label{fig:ULmcDM2p}
 \end{center}
\end{figure}

In the top panel of Fig.~\ref{fig:ULmcDM2p}, the upper limits are
plotted for the case of the $2\%$ milli-charged DM. 
The black dashed line is the cross section corresponding to the DM model
that could explain the EDGES results. 
The SK limit is not strong enough to rule out the milli-charged DM
model. 
Based on 3000 running days, the predictions show that Hyper-Kamiokande,
JUNO and DUNE cannot reach the desired limit as well. 
However, the actual data will probably induce some fluctuations,
possibly resulting in a stronger limit, as can be seen from the behavior
of the Monte Carlo region of Hyper-Kamiokande, compared to its predicted
line. 
The strongest limit comes from DUNE. 
Running the detector long enough might result in strong enough limits to
constrain the milli-charged DM model. A combined analysis of the data of the several experiments could result in a stronger limit by up to a factor of 2. 
In the bottom panel of Fig.~\ref{fig:ULmcDM2p}, we show the limits in
terms of the coupling constants and the masses of the dark matter
particle and the new mediator, $(g_{\chi}g_f)^2(m_{\chi}m_V)^4$,
specific to our DM model.

\begin{figure}
\includegraphics[width=7cm]{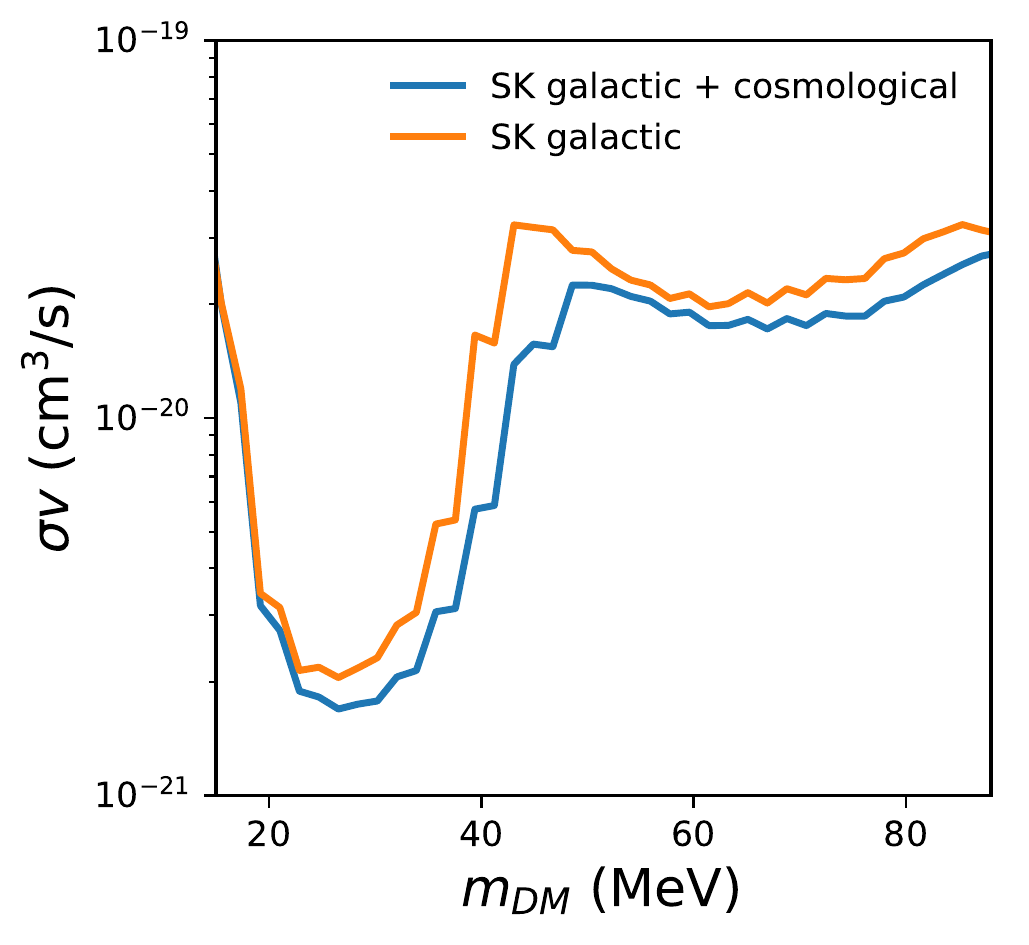}
\caption{Upper limits on the annihilation cross section of milli-charged
 Dirac DM, making up 2$\%$ of the total DM, as a function of the DM
 mass, computed from the analysis of 2853 days of SK data. The orange
 line only contains the Galactic contribution to the neutrino flux
 coming from DM annihilations, while the blue line also includes the
 extragalactic contribution.}\label{fig:SKULcompare}
\end{figure}

Figure~\ref{fig:SKULcompare} shows the SK limits both with and without
taking the extragalactic DM annihilation flux into account. 
We note that the Galactic flux has a substantial contribution to the
limit.

We note that the most recent CMB (re)analysis find that the fraction of
milli-charged DM might be contrained even more tightly, $f_\chi \lesssim
0.4$\%~\cite{Boddy:2018wzy} (and references therein).
In the case of $f_\chi = 0.4$\%, our limits get weaker by a factor of
$(2/0.4)^2 = 25$ [Eq.~(\ref{eq:flux})].
However, the annihilation cross section required to explain the relic
abundance becomes larger by a factor of $2/0.4 = 5$
[see Eq.~(\ref{eq:sigmav}) and subsequent sentences].
Hence our limits on the annihilation cross section relative to its
canonical value will be weakened by a factor of 5.

\begin{figure}[h!]
 \begin{center}
  \includegraphics[width=8.5cm]{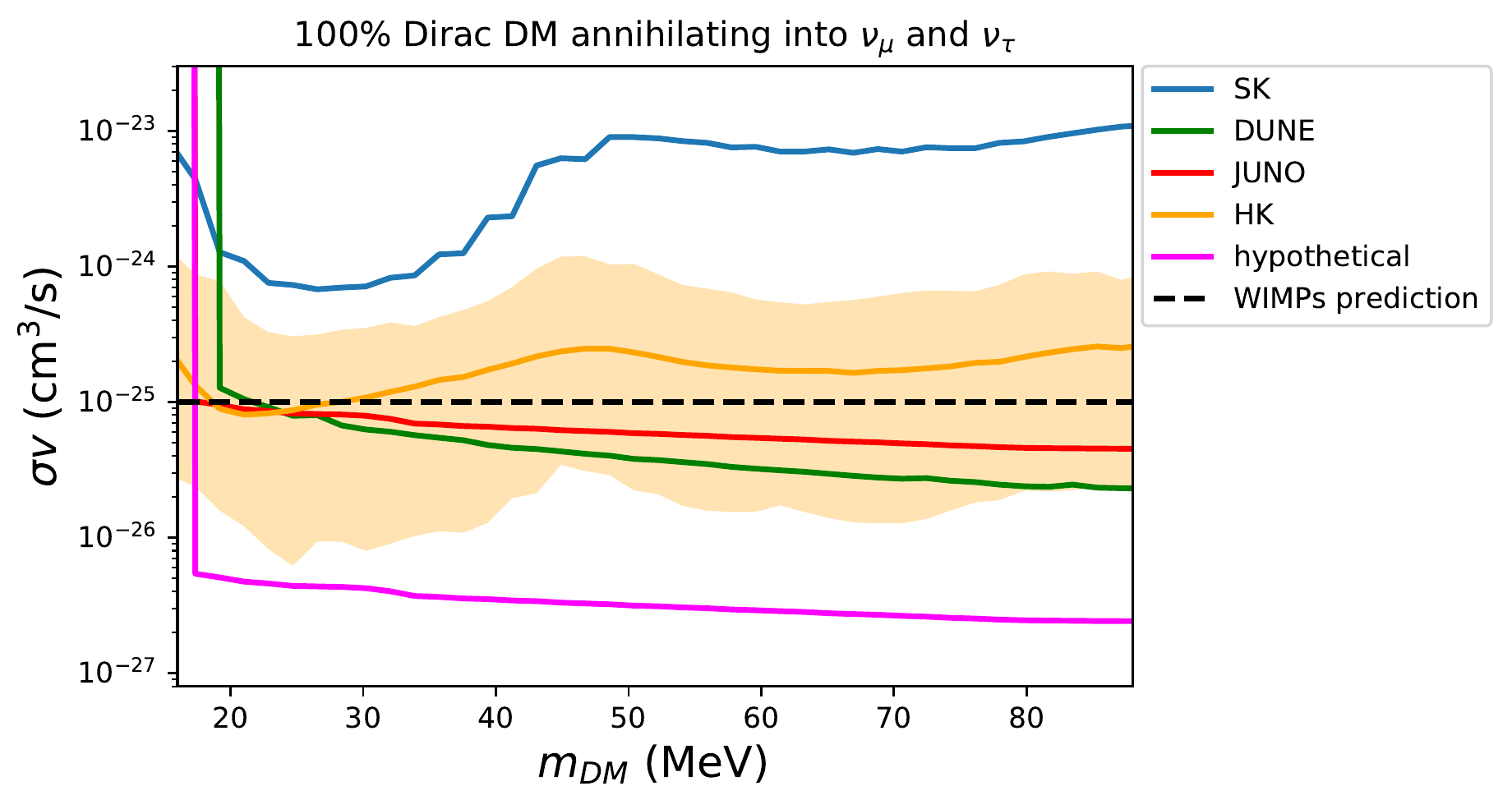} 
  \includegraphics[width=8.5cm]{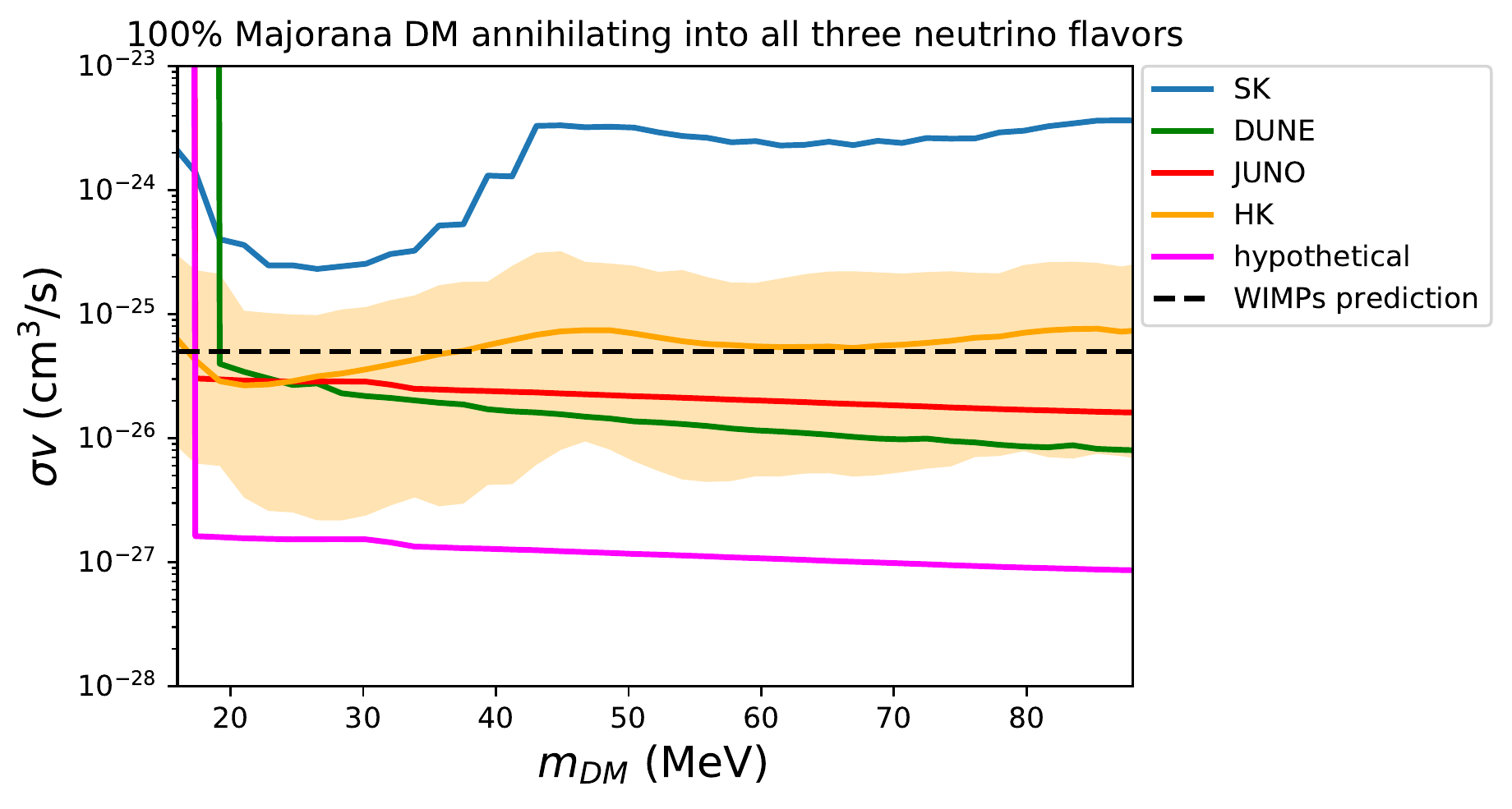}
  \caption{Upper limits on the annihilation cross section of
  milli-charged Majorana DM, being the total 100$\%$ of DM, as a
  function of the DM mass. The upper panel shows the case of Dirac DM
  annihilating into only muon and tau neutrinos. The lower panel shows
  the case of Majorana DM annihilating into all three neutrino flavors
  with equal fraction. The black dashed line is the preferred cross
  section for WIMPs, in case of Dirac (Majorana) DM in the upper (lower)
  panel. The blue line comes from the analysis of 2853 days of SK
  data. The other lines are predictions using
  Swordfish~\cite{Edwards:2017kqw, Edwards:2017mnf}. The orange band
  shows the region between the minimum and maximum upper limit values
  for Hyper-Kamiokande predicted through a Monte Carlo
  simulation.}\label{fig:ULmcDM100p}
 \end{center}
\end{figure} 

Finally, besides exploring the milli-charged DM model, we explore two
more generic cases, where 100\% of DM annihilates into neutrinos.
First, we study the same Dirac DM annihilating into only muon and tau
neutrinos, whose results are shown in the top panel of
Fig.~\ref{fig:ULmcDM100p}.
The second case is Majorana DM annihilating into 3 neutrino flavors with equal
fraction, shown in the bottom panel of Fig.~\ref{fig:ULmcDM100p}.
Since neutrinos are harder to detect than gamma rays, this results in a
more conservative, and hence most general constraints on the DM
annihilation~\cite{Yuksel:2007ac}.
While the current limit of SK could not yet constrain the WIMP
prediction, in both of these cases, Hyper-Kamiokande, JUNO and DUNE will
certainly be able to do so. 
We note that our updated limit is weaker than the limit found in
Ref.~\cite{PalomaresRuiz:2007eu} based on the previous data set of
SK~\cite{Malek:2002ns} by a factor of several.

\section{Conclusions}\label{sec:conclusion}

The recent results of EDGES suggest that the primordial gas underwent
extensive cooling from some additional DM kind. 
Several constraints on the DM parameter space have already been made. 
We explore the yet unconstrained region in the case that milli-charged
DM makes up for $\sim$2\% of the total dark matter, through the scenario
in which this DM interacts with the standard model through the $\mu-\tau$ lepton
number.
This additional interaction is motivated by the thermal freeze-out
scenario to explain the correct relic density, and also by the fact that
it is largely unconstrained.
If this DM has masses of 10--100~MeV as suggested by the EDGES measurement,
it annihilates only into mu and tau neutrinos.

By calculating the neutrino flux from the Galactic and extragalactic
halos and comparing with existing data, we find that data from
Super-Kamiokande are not yet able to constrain this model.
We however find that future experiments might be able to detect
neutrinos from this particular DM species.
The hypothetical experiment that we study with the size comparable to
Hyper-Kamiokande and energy resolution comparable to JUNO or DUNE would
be able to reach the desired limits. 
Although such an experiment is not scheduled to be build in the near
future, there has been a European-wide initiative to study the
possibility of an experiment with a size of the right order of
magnitude~\cite{Angus:2010sz}.
We furthermore provide updated limits on the annihilation cross section
for more general WIMP DM model in the mass range of 16--88~MeV, using
(expected) data from the current and future neutrino experiments. 
We find that the current data of SK can not yet put constraints on the
WIMPs prediction, but the future experiments Hyper-Kamiokande, DUNE and
JUNO will be capable of this.

\acknowledgments
We thank Thomas Edwards for the useful discussions. This work was supported by the Foundation for Fundamental Research on Matter (FOM) through the FOM Program (N.K. and S.A.), and partly financed by JSPS KAKENHI Grant Numbers JP17H04836, JP18H04340, and JP18H04578 (S.A.).

% Bibliography
\bibliographystyle{h-physrev}
\bibliography{mybib}

\end{document}